\documentclass{elsart}

\usepackage{enumerate}
\usepackage[latin1]{inputenc}
\usepackage[T1]{fontenc}
\usepackage{amssymb,amsmath}
\usepackage[final]{graphicx}
\usepackage{xspace}
\usepackage{rotating}
\usepackage{color}

\newtheorem{theorem}{Theorem}[section]
\newtheorem{corollary}[theorem]{Corollary}
\newtheorem{definition}[theorem]{Definition}

\newtheorem{lemma}[theorem]{Lemma}

\renewcommand{\Box}{{\vrule width1.2ex height1.2ex depth0cm}}
\newenvironment{proof}{\medskip \noindent{\bf Proof }}{\hfill \Box \medskip\medskip}

\newcommand{\map}[3]{{#1} : {#2} \rightarrow {#3}} 

\newcommand{\concat}[1]{\langle {#1} \rangle}

\newcommand{\ram}{\mbox{\rm RAM}}

\newcommand{\np}{\mathrm{NP}}
\newcommand{\p}{\mathrm{P}}

\newcommand{\dtime}{\mathrm{DTIME}}
\newcommand{\ntime}{\mathrm{NTIME}}
\newcommand{\lin}{\mathrm{LIN}}
\newcommand{\tmlin}{\mathrm{TM-LIN}}
\newcommand{\turingm}{\mathrm{TM}}
\newcommand{\nlin}{\mathrm{NLIN}}
\newcommand{\dlin}{\mathrm{DLIN}}

\newcommand{\accept}{accept}
\newcommand{\reject}{reject}

\newcommand{\sat}{\textsc{sat}}

\newcommand{\risa}{\textsc{risa}}

\DeclareSymbolFont{AMSb}{U}{msb}{m}{n}
\DeclareMathSymbol{\N}{\mathbin}{AMSb}{"4E}
\DeclareMathSymbol{\F}{\mathbin}{AMSb}{"46}
\DeclareMathSymbol{\Z}{\mathbin}{AMSb}{"5A}
\DeclareMathSymbol{\R}{\mathbin}{AMSb}{"52}
\DeclareMathSymbol{\Q}{\mathbin}{AMSb}{"51}
\DeclareMathSymbol{\I}{\mathbin}{AMSb}{"49}
\DeclareMathSymbol{\C}{\mathbin}{AMSb}{"43}
\DeclareMathSymbol{\BL}{\mathbin}{AMSb}{"4C}

\newcommand{\card}[1]{\left|#1\right|} 

\begin{document}
\title{On the structure of linear-time reducibility}
\author{Philippe Chapdelaine}
\date{}

\maketitle

\begin{abstract}
In 1975, Ladner showed that under the hypothesis 
$\p \neq \np$, there exists a language which is
neither in $\p$, nor $\np$-complete. 
This result was latter generalized by Schöning 
and several authors to various 
polynomial-time complexity classes.
We show here that such results also apply to
linear-time reductions on \ram s (resp.
Turing machines), and hence allow 
for separation results in linear-time classes
similar to Ladner's ones for polynomial time.
\end{abstract}

\section{Introduction}

$\dlin$ is the class of the (decision) problems decided by 
{\em deterministic} \ram s in time $O(n)$.
Likewise $\nlin$ is the class of problems
decided by  {\em nondeterministic} \ram s in time 
$O(n)$.
These classes formalize the intuitive notion of
algorithms working in linear 
time \cite{grandjean94,grandjean96sorting,Schwentick97,GS02}.

One of the features of $\nlin$ that makes it quite interesting 
to study is that it contains
most of the {\em natural} $\np$-complete problems, among them
the 21 problems from \cite{karp72}.
Also, the problem $\risa$ (``Reduction of Incompletely
Specified finite Automaton'', Problem AL7 in \cite{gareyjohnson}) 
has been shown to be complete
in $\nlin$ under linear-time computable reductions \cite{grandjean90}.
Moreover, a series of articles have shown 
the robustness of both these classes, 
through logical, algebraic or computational means
(see for example \cite{Schwentick97,GS02,GO03}).
So, the conjecture $\nlin \neq \dlin$, a weaker version
of $\np \neq \p$, appears quite central in the study
of the complexity of natural $\np$-complete problems.

On the other hand, it is well known that natural problems in $\np$
are proved to be either in $\p$ or $\np$-complete, except for a very small number
of them among which the most prominent is the Graph Isomorphism problem.
In fact, Ladner proved in \cite{ladner75} that under the conjecture
$\p \neq \np$, there exist problems in $\np \smallsetminus \p$
that are not $\np$-complete.
This result has since been generalized with the Uniform Diagonalization
method \cite{schoning82,BD82} that applies to many other complexity classes,
as well as other notions of reductions \cite{reganvollmer97,regan83}.

However, whereas numerous results were obtained for various polynomial-time
and polynomial-space complexity classes, with the appropriate reductions,
no similar result has ever been proved for linear-time complexity classes
and reductions, until now.
Such results may prove significant, considering the following facts. 
\begin{enumerate}
 \item Most {\em natural} $\np$-complete problems belong to $\nlin$.
 \item In contrast with  $\np$, there seem to be {\em very few} 
$\nlin$-complete problems, such as $\risa$, 
and many intermediary problems. In fact, the archetypical 
$\np$-complete problem $\sat$
doesn't seem to be $\nlin$-complete, as it uses a 
sublinear number of nondeterministic instructions, namely $O(n / \log n)$
(see \cite{grandjean94}), and {\em a lot} of natural 
$\np$-complete problems are {\em linearly} equivalent to $\sat$:
e.g. {\sc Vertex Cover}, {\sc Dominating Set}, 
$3$-{\sc Colorability}, etc., as shown in 
\cite{BarbanchonG02,creignou95lin,dewdney,grandjean96sorting}.
\end{enumerate}

In this note, we strengthen Ladner's and Balc\'azar and D\'iaz's 
results \cite{ladner75,BD82} for polynomial-time degrees 
by proving that they similarly hold for 
\ram s (resp. Turing Machines) linear-time degrees.
We obtain these results by essentially noticing that
the concepts and proofs of \cite{BDG1} and \cite{papadimitriou94} 
for the polynomial case (attributed to \cite{ladner75,LLR81,schoning82})
work or can be adapted in the linear case.
For example, we deduce from the separation result 
$\dtime_{\turingm}(n) \subsetneq \ntime_{\turingm}(n)$
(by \cite{ppst83}) on Turing machines that there exists 
an infinite number of pairwise incomparable problems which 
are neither in $\dtime_{\turingm}(n)$ nor $\nlin$-complete
under linear reductions on Turing machines. Note that
this result holds {\em without any hypothesis}.

\section{The Uniform Diagonalization Theorem for linear time}

We first give some definitions and preliminary results.
The computation model used is the \ram\ model, as it was defined
in \cite{Schwentick97,GS02}. That is with a unary 
structure $w=([n],f)$,
$[n]= \{0,1,\ldots,n-1\}$ and $\map{f}{[n]}{[n]}$,
as input\footnote{We write $n=\card w$ and call it the {\em size} of the input.}
and with a specified set of allowed (classical) instructions.
A ({\em decision}) {\em problem}, also called {\em set} or {\em language}, 
is a set of input structures.
A \ram\ works in {\em linear time} if for each input (structure) $w$
of size $n$ it performs $O(n)$ instructions and uses
only integers $O(n)$ (as register contents and addresses).
$\dlin$ (resp. $\nlin$) is the class of problems decided
by deterministic (resp. nondeterministic) \ram s in linear time.
(Note that it was shown, see for example \cite{GS02}, 
that the linear-time classes are quite robust
and are essentially independent of the set of allowed instructions.)

\begin{definition}
A class ${\mathcal C}$ of recursive sets is {\em recursively presentable} if there
exists an effective enumeration $M_1,M_2,\ldots$ of deterministic
\ram s which halt on all their inputs, and such that 
${\mathcal C}=\{ L(M_i) \ |\ i=1,2,\ldots \}$.
\end{definition}

By convention, the empty class is recursively presentable.

It is easy to see that $\dlin$ is 
recursively presentable. In fact, one can check that every pair $(M,c)$, 
where $M$ is a deterministic \ram\ $M$ and $c$ is an integer,
defines  
$L_c(M)= \{ w \ |\ M \text{ accepts $w$ in time at most } c \card w \}$, 
which is in $\dlin$, and conversely, every language in $\dlin$ is of
this form. So any effective enumeration of all the pairs $(M,c)$ is a recursive
presentation of $\dlin$.

\begin{definition}
A class of sets ${\mathcal C}$ is {\em closed under finite variants}
if, for every $A,B$ such that $A \in {\mathcal C}$ 
and the symmetric difference $A \Delta B$ is finite,
we have $B \in {\mathcal C}$.
\end{definition}

We can  now  prove that  the Uniform Diagonalization Theorem,
first given by Schöning \cite{schoning82} (see also \cite[Theorem 7.4]{BDG1}
and \cite{fortnowDiag})
for polynomial-time computable reductions,
can be strengthen to apply to linear-time reductions on \ram s, denoted $\leq_\lin$ 
(we write $A \leq_\lin B$ for two problems $A$ and $B$
to mean that there is some linear-time many-one reduction
from $A$ to $B$ that is computable on some \ram).
A {\em linear degree} is an equivalence class of some problem $A$:
$\{ B\ |\ A \leq_\lin B \text{ and } B \leq_\lin A \}$.

\begin{theorem} \label{UDT}
Let ${\mathcal C}_1$ and ${\mathcal C}_2$ be two recursively presentable classes
(of recursive sets),
both closed under finite variants. Let $A_1$ and $A_2$ be 
two recursive sets such that $A_1 \notin {\mathcal C}_1$ and
$A_2 \notin {\mathcal C}_2$. 
Then there exists a set $A$ such that $A \notin {\mathcal C}_1$,
$A \notin {\mathcal C}_2$ and $A \leq_\lin A_1 \oplus A_2$.
\end{theorem}

Here, $A_1 \oplus A_2$ denotes the disjoint union of $A_1$ and $A_2$, 
that is 
$\{ \concat{w,0} \ |\ w \in A_1 \} \cup \{ \concat{w,1} \ |\ w \in A_2 \}$,
where $\concat{x,y}$ is any reversible pairing operation
computable in linear time.

The proof given here is a mixture of the one given for Schöning's result 
for polynomial reductions as
it appears in \cite{BDG1}, which does not seem to apply to linear reductions,
and the one given for the famous Ladner's Theorem in
\cite{papadimitriou94}, which implicitly applies to linear reductions. 
Here, ``almost always'' will stand for ``except for finitely many cases''.

\begin{proof}
Let $M_0^1,M_1^1,\ldots$ be a recursive presentation of
${\mathcal C}_1$, and $M_0^2,M_1^2,\ldots$ be a recursive presentation of
${\mathcal C}_2$. Let $S_1$ be a \ram\ that decides $A_1$ and
$S_2$ be one that decides $A_2$.

The set $A$ will be the following one:
$$
A=    \bigl( A_1 \cap \{ x \ |\ f(\card x)\text{ is even}  \} \bigr) 
\cup  \bigl( A_2 \cap \{ x \ |\ f(\card x)\text{ is odd}  \} \bigr), 
$$
where function $f$ will be such that if $A \in {\mathcal C}_1$ 
then $f(n)$ is almost always even, and if $A \in {\mathcal C}_2$ 
then $f(n)$ is almost always odd.
So, if $A \in {\mathcal C}_1$, then $A$ is almost always equal to $A_1$. 
Given that ${\mathcal C}_1$ is closed under finite variants, 
this proves that $A_1 \in {\mathcal C}_1$, in contradiction with the original
hypothesis. A similar reasoning applies if $A \in {\mathcal C}_2$.

The definition of function $f$, or more precisely of the \ram\ $F$
that computes it, is given by a recursion scheme that defines
along the \ram\ $K$ that recognizes~$A$.

The \ram\ $F$ takes as input an integer $n \in \N$ and
computes $f(n)$. If $n=0$, then $F$ outputs $1$ (that is $f(0)=1$).
Otherwise, $F$ first recursively computes as many values 
$f(0), f(1), f(2), \ldots$, as 
it is able to complete in exactly $n$ steps.
Suppose that the last value $i$ for which it is possible
to complete the computation is $f(i)=k$. 
Then $F$ proceeds in two different ways,
depending on whether $k$ is even or odd.

\begin{enumerate}
 \item If $k=2j$ is even, then $F$ starts computing 
$M_j^1(z), S_1(z),S_2(z)$ and $F(\card z)$, where $z$ ranges
lexicographically over every possible input structure of
size $1,2,\ldots$, as many as it is possible to complete in $n$ steps
of computation.
Its aim is to find a structure $z$ such that
$K(z) \neq M_j^1(z)$,
that is a $z$ that verifies one of the following conditions:
\begin{enumerate}[(a)]
 \item $M_j^1(z)=\accept$, $f(\card z)$ is odd, and $S_2(z)=\reject$;
 \item $M_j^1(z)=\accept$, $f(\card z)$ is even, and $S_1(z)=\reject$;
 \item $M_j^1(z)=\reject$, $f(\card z)$ is odd, and $S_2(z)=\accept$;
 \item $M_j^1(z)=\reject$, $f(\card z)$ is even, and $S_1(z)=\accept$;
\end{enumerate}
If such a $z$ can be found in $n$ steps, then $f(n)=k+1$,
otherwise $f(n)=k$.

\item If $k=2j+1$ is odd, then do as above, but with $M_j^2$ instead 
of $M_j^1$, trying to find a $z$ such that
$K(z) \neq M_j^2(z)$.
Again, if such a $z$ is found, then  $f(n)=k+1$,
otherwise $f(n)=k$.
\end{enumerate}
Note that on input $n$, $F$ works in {\em exactly} $2n$ steps.

It is easy to show, recursively, that $f$ is a non-decreasing function,
whose set values consists of the consecutive integers $1,2,3,\ldots$.
We now show that $f$ is not bounded. This will imply that
$A$ is neither in ${\mathcal C}_1$ nor in ${\mathcal C}_2$, 
as there will be no $M_i^1$ 
(resp. $M_i^2$) such that $K$ and $M_i^1$ (resp. $M_i^2$) 
decide the same language.

Suppose that there exist $n_0$ and $p$ such that $f(n)=2p$ for every
$n \geq n_0$. This means that for each $z$, $K(z) = M_p^1(z)$, 
and hence $A \in {\mathcal C}_1$. But then,
this also means that $f$ is even and so $A$ is almost always equal to $A_1$. 
Since ${\mathcal C}_1$ is 
closed under finite variants, we deduce that $A_1 \in {\mathcal C}_1$, 
in contradiction
with the  original hypothesis. A similar reasoning holds if
there exist  $n_0$ and $p$ such that $f(n)=2p+1$ for every
$n \geq n_0$.

Now, there remains to prove that $A \leq_\lin A_1 \oplus A_2$.
Given an input $x$, suppose that $f(\card x)$ is even
(resp. odd), then the following equivalences hold:
\begin{eqnarray*}
x \in A         &  \Leftrightarrow x \in A_1 &
        \Leftrightarrow \concat{x,0} \in A_1 \oplus A_2 \\
\bigl( \text{resp. }
x \in A \bigr. & \Leftrightarrow x \in A_2  & \bigl. 
        \Leftrightarrow \concat{x,1} \in A_1 \oplus A_2
\bigr).
\end{eqnarray*}
This shows that the transformation
$$
R(x)=\left\{
\begin{tabular}{ll}
$\concat{x,0}$ & if $f(\card x)$ is even \\
$\concat{x,1}$ & if $f(\card x)$ is odd,
\end{tabular}
\right.
$$
which is computable in linear time,
is a reduction from $A$ to $A_1 \oplus A_2$.
\end{proof}

\section{The structure of nondeterministic linear time}

The following lemmas show that the linear-time classes
are recursively presentable.
The proofs of Lemma \ref{reduction} and Lemma \ref{two results} 
are similar to the proofs of \cite[Lemma 7.5]{BDG1}
and \cite[Lemma 7.7]{BDG1} respectively. One need only notice
that the reasoning is still true with linear-time computable reductions,
and apply the lemmas with the good parameters.

\begin{lemma} \label{reduction}
The class of the languages reducible to $\risa$ 
(Reduction of Incompletely Specified Automaton)
in linear time, 
that is $\nlin$, is recursively presentable.
\end{lemma}

\begin{lemma} \label{two results}
The class of the $\nlin$-complete problems is recursively presentable.
\end{lemma}

We can now apply the Uniform Diagonalization Theorem 
to obtain a result on the structure of the nondeterministic linear-time
class $\nlin$,
similar to the one obtained by Ladner for $\np$.

\begin{theorem}
If $\dlin \subsetneq \nlin$, then there exists
a language in $\nlin$ which is neither in $\dlin$, nor
$\nlin$-complete.
\end{theorem}

\begin{proof}
Apply Theorem \ref{UDT} with the following parameters:
${\mathcal C}_1=\dlin$, ${\mathcal C}_2$ is the class of the $\nlin$-complete
problems, $A_1=\risa$, and $A_2=\emptyset$.
\end{proof}

As it is the case for the class $\np$, it is even possible, under some
similar hypothesis, to prove that, not only there exist intermediate problems,
but also that there are an infinite number of pairwise incomparable 
(through linear-time reductions) problems.
The following theorem is proved by noticing that 
the proof of \cite[Theorem 7.10]{BDG1} also
applies to linear-time reductions.

\begin{theorem} \label{infinite}
Let $A$ and $B$ be two recursive languages such that
$A \leq_\lin B$ but $B \not \leq_\lin A$.
Then, there exists an infinite family of languages 
$D_i$, $i \in \N$, such that:
\begin{enumerate}[(a)]
 \item for all $i$, $A \leq_\lin D_i \leq_\lin B$, but 
$B \not\leq_\lin D_i \not\leq_\lin A$;
 \item for all $i,j$, if $i \neq j$ then $D_i \not \leq_\lin D_j$
and $D_j \not \leq_\lin D_i$.
\end{enumerate}
\end{theorem}

Applying this last theorem to classes $\dlin$ and $\nlin$,
with $A = \emptyset$ and $B= \risa$,
we get the following corollary.

\begin{corollary}
If $\dlin \subsetneq \nlin$, then there exist
infinitely many pairwise incomparable linear degrees
between $\dlin$ and the class of the $\nlin$-complete problems.
\end{corollary}

An interesting feature of these results is that they are still
true if we consider linear-time reductions on {\em Turing Machines},
denoted $\leq_\tmlin$, 
rather than those on \ram s, which were denoted $\leq_\lin$. 
The former reductions are more precise (i.e. restricted)
than the latter but the known $\nlin$-complete problems
(under linear-time reductions on \ram s), 
typically $\risa$, remain $\nlin$-complete
under linear-time reductions on Turing Machines\footnote{The 
more general question of whether the two notions of
$\nlin$-completeness for linear-time reductions on \ram s
or on Turing machines are equivalent is an open problem.}.
Now, consider the following inclusions:
$$
\dtime_\turingm(n) \subsetneq \ntime_\turingm(n) \subseteq \nlin,
$$
which were proved in \cite{ppst83} and \cite{grandjean90} respectively, where
$\dtime_\turingm(n)$ (resp. $\ntime_\turingm(n)$) is the class
of the problems computable in linear time on
deterministic Turing Machines (resp. nondeterministic Turing Machines).
We can thus deduce the following theorem, which does not require
{\em any} hypothesis.

\begin{theorem}
There exists a problem (in fact, an infinite number of
pairwise incomparable problems) in $\nlin$, which is neither
in $\dtime_\turingm(n)$, nor $\nlin$-complete
under linear-time reductions on Turing Machines.
\end{theorem}

\section{Conclusion}

This note shows that linear-time reducibility $\leq_\lin$
(on the \ram\ model),
a much more precise notion than the usual
polynomial-time reducibility, shares the same properties as this last one.
We present the first structural
complexity results for linear-time complexity classes
and linear-time reducibility.
Another interesting point is that our results can be similarly applied
to the $\sat$ linear degree, i.e., the 
class of the (many) problems linearly equivalent to $\sat$
(under $\leq_\lin$-reductions),
a problem which is conjectured not to be $\nlin$-complete:
we again obtain infinitely many pairwise incomparable linear degrees
on the one hand between the $\sat$ degree and $\dlin$ (if $\sat \notin \dlin$), 
and on the other hand between the $\sat$ degree and the class 
of the $\nlin$-complete problems
(if $\sat$ is not $\nlin$-complete, which is a reasonable conjecture).
We also show, {\em without any hypothesis}, the existence of infinitely many 
pairwise incomparable linear degrees between 
the problems computable in linear time
on {\em deterministic Turing Machines} and the class of the
problems in $\nlin$ which are $\nlin$-hard 
under linear-time reductions on Turing Machines.
Finally, we believe that these results give arguments for the robustness 
and significance of linear-time
reductions and linear degrees, either on the \ram\ 
model or the Turing model.

\bibliographystyle{plain}
\bibliography{bibliographie}

\begin{thebibliography}{10}

\bibitem{BD82}
J.~L. Balc\'{a}zar and J.~D\'{i}az.
\newblock A note on a theorem by {L}adner.
\newblock {\em Information Processing Letters}, 15:84--86, 1982.

\bibitem{BDG1}
J.~L. Balc\'{a}zar, J.~D\'{i}az, and J.~Gabarr\'{o}.
\newblock {\em Structural Complexity {I}}.
\newblock EATCS Monographs on Theoretical Computer Science. Springer-Verlag,
  1988.

\bibitem{BarbanchonG02}
R.~Barbanchon and E.~Grandjean.
\newblock Local problems, planar local problems and linear time.
\newblock In {\em Proceedings of CSL 2002}, pages 397--411. Springer, 2002.

\bibitem{creignou95lin}
N.~Creignou.
\newblock The class of problems that are linearly equivalent to satisfiability
  or a uniform method for proving {NP}-completeness.
\newblock {\em Theoretical Computer Science}, 145(1/2):111--145, 1995.

\bibitem{dewdney}
A.K. Dewdney.
\newblock Linear time transformations between combinatorial problems.
\newblock {\em International Journal of Computer Mathematics}, 11(2):91--110,
  1982.

\bibitem{fortnowDiag}
L.~Fortnow.
\newblock {\em Current Trends in Theoretical Computer Science, Entering the
  21th Century}, chapter Diagonalization.
\newblock World Scientific, 2001.

\bibitem{gareyjohnson}
M.~R. Garey and D.~S. Johnson.
\newblock {\em Computers and intractability: A guide to the theory of
  {NP}-completeness}.
\newblock W. H. Freeman and Company, 1979.

\bibitem{grandjean90}
E.~Grandjean.
\newblock A nontrivial lower bound for an {NP} problem on automata.
\newblock {\em SIAM Journal on Computing}, 19(3):438--451, 1990.

\bibitem{grandjean94}
E.~Grandjean.
\newblock Linear time algorithms and {NP}-complete problems.
\newblock {\em SIAM Journal on Computing}, 23(3):573--597, 1994.

\bibitem{grandjean96sorting}
E.~Grandjean.
\newblock Sorting, linear time and the satisfiability problem.
\newblock {\em Annals of Mathematics and Artificial Intelligence}, 16:183--236,
  1996.

\bibitem{GO03}
E.~Grandjean and F.~Olive.
\newblock Graph properties checkable in linear time in the number of vertices.
\newblock {\em Journal of Computer and System Sciences}, 68(3):546--597, 2004.

\bibitem{GS02}
E.~Grandjean and T.~Schwentick.
\newblock Machine-independent characterizations and complete problems for
  deterministic linear time.
\newblock {\em SIAM Journal on Computing}, 32(1):196--230, 2002.

\bibitem{karp72}
R.~M. Karp.
\newblock Reducibility among combinatorial problems.
\newblock In R.~E. Miller and J.~W. Thatcher, editors, {\em Complexity of
  Computer Computations}, pages 85--103. Plenum Press, New York, 1972.

\bibitem{ladner75}
R.~Ladner.
\newblock On the structure of polynomial-time reducibility.
\newblock {\em Journal of the ACM}, 22:155--171, 1975.

\bibitem{LLR81}
L.~Landweber, R.~Lipton, and E.~Robertson.
\newblock On the structure of sets in {NP} and other complexity classes.
\newblock {\em Theoretical Computer Science}, 15:181--200, 1981.

\bibitem{papadimitriou94}
C.~H. Papadimitriou.
\newblock {\em Computational Complexity}.
\newblock Addison-Wesley Publishing Company, 1994.

\bibitem{ppst83}
W.~J. Paul, N.~Pippenger, E.~Szemer{\'e}di, and W.~T. Trotter.
\newblock On determinism versus non-determinism and related problems.
\newblock In {\em Proceedings of the 24th IEEE Symposium on Foundations of
  Computer Science}, pages 429--438, 1983.

\bibitem{regan83}
K.~W. Regan.
\newblock On diagonalization methods and the structure of language classes.
\newblock In {\em Fundamentals of Computation Theory}, Lecture Notes in
  Computer Science, pages 368--380, 1983.

\bibitem{reganvollmer97}
K.~W. Regan and H.~Vollmer.
\newblock Gap-languages and log-time complexity classes.
\newblock {\em Theoretical Computer Science}, 188:101--116, 1997.

\bibitem{schoning82}
U.~Schöning.
\newblock A uniform approach to obtain diagonal sets in complexity classes.
\newblock {\em Theoretical Computer Science}, 18:95--103, 1982.

\bibitem{Schwentick97}
T.~Schwentick.
\newblock Algebraic and logical characterizations of deterministic linear time
  classes.
\newblock In {\em Proceedings of the 14th Symposium on Theoretical Aspects of
  Computer Science STACS 97}, pages 463--474, 1997.

\end{thebibliography}

\end{document}